\documentclass[traditabstract,letter]{aa} 
\usepackage{graphicx}
\usepackage{txfonts}
\newcommand{\Rin}{R_{\rm in}}
\newcommand{\Rout}{R_{\rm out}}
\newcommand{\Tsub}{T_{\rm sub}}

\newcommand{\Tsf}{T_{\rm max}}

\usepackage{natbib}
\begin{document}
\title{Possible evidence for a common radial structure in nearby AGN tori}


\author{Makoto Kishimoto,
       Sebastian F. H\"onig,
       Konrad R.~W.~Tristram
       \and
       Gerd Weigelt
       }

\institute{
           Max-Planck-Institut f\"ur Radioastronomie, Auf dem H\"ugel 69,
           53121 Bonn, Germany\\
           \email{mk@mpifr-bonn.mpg.de}
          }

   \date{Received 30 September 2008; accepted 10 December 2008}

\authorrunning{Kishimoto et al.}

\titlerunning{A common radial structure in AGN tori}

 
\abstract{ We present a quantitative and relatively model-independent
way to assess the radial structure of nearby AGN tori.  These putative
tori have been studied with long-baseline infrared (IR)
interferometry, but the spatial scales probed are different for
different objects.  They are at various distances and also have
different physical sizes which apparently scale with the luminosity of
the central engine.  Here we look at interferometric visibilities as a
function of spatial scales normalized by the size of the inner torus
radius $\Rin$.  This approximately eliminates luminosity and distance
dependence and, thus, provides a way to uniformly view the
visibilities observed for various objects and at different
wavelengths.  We can construct a composite visibility curve over a
large range of spatial scales if different tori share a common radial
structure.  The currently available observations do suggest
model-independently a common radial surface brightness distribution in
the mid-IR that is roughly of a power-law form $r^{-2}$ as a function
of radius $r$, and extends to $\sim$100 times $\Rin$.  Taking into
account the temperature decrease toward outer radii with a simple
torus model, this corresponds to the radial surface density
distribution of dusty material directly illuminated by the central
engine roughly in the range between $r^0$ and $r^{-1}$.  This should
be tested with further data.  }

\keywords{Galaxies: active, Galaxies: Seyfert, Infrared: galaxies, 
Techniques: interferometric}

\maketitle
%

\section{Introduction}

Speckle and long-baseline interferometry in the infrared (IR) has
started to explore the spatial structure of the putative tori in
Active Galactic Nuclei \citep[AGN; e.g.][]{Wittkowski98,Weinberger99,
Swain03,Weigelt04,Jaffe04,Wittkowski04,Meisenheimer07,Tristram07}.
However, the number of long baselines per object is still generally
limited, and a given baseline probes different spatial scales for
targets at various distances and having different physical sizes.
Here we investigate the possibility of uniformly view these
interferometric measurements taken for various objects spanning over
different wavelengths, by studying them as a function of spatial
scales in units of torus inner radius.  


\begin{table*}
\caption[]{List of AGNs used in this paper.}
\begin{tabular}{llcccccccccc}
\hline
name    & type & $cz^a$       & obs. date & inst. & $\lambda$  & $b$ & PA       & $\Rin$ & K-band     & reference & $\Delta$PA$^g$\\ 
        &      & (km s$^{-1}$)& (UT)      &       & ($\mu$m)   & (m) & ($\deg$) & (mas)  & Lag (days) &           & (deg)         \\ 
\hline
NGC1068  & 2 &  914 & 2003-06-15/16 & MIDI & 8.2-13 & 79 &-179 &  1.6$^b$ &        & \cite{Jaffe04}   & 89\\ 
NGC1068  & 2 &  914 & 2003-11-09/10 & MIDI & 8.2-13 & 46 &-135 &  1.6$^b$ &        & \cite{Jaffe04}   & 45\\ 
Mrk1239  & 1 & 6321 & 2005-12-19    & MIDI & 8.2-13 & 41 &  36 & 0.17$^c$ &        & \cite{Tristram08}& 86\\ 
NGC3783  & 1 & 3234 & 2005-05-28    & MIDI & 8.2-13 & 43 &  46 & 0.32     & 85$^d$ & archive          &  0\\ 
NGC3783  & 1 & 3234 & 2005-05-31    & MIDI & 8.2-13 & 65 & 120 & 0.32     & 85$^d$ & \cite{Beckert08} & 74\\ 
NGC4151  & 1 & 1242 & 2003-05-23    & KI   & 2.2    & 83 &  37 & 0.47     & 48$^e$ & \cite{Swain03}   & 50\\ 
\hline
\end{tabular}
\\
$^a$Radial velocity corrected for the cosmic microwave background from NED.\\
$^b$Calculated from the estimated intrinsic $V$-band flux 
\citep{Pier94} and Eq.\ref{eq-Rin}.\\
$^c$Calculated from the $V$-band flux $f_{\nu}=7.8$mJy \citep{Veron03} and Eq.\ref{eq-Rin}.\\
$^d$\cite{Glass92}. $^e$\cite{Minezaki04}. $^f$\cite{Suganuma06}.\\
$^g$Position angle difference between the projected baseline and
the expected torus major axis direction
(see text).

\label{tab-obj}
\end{table*}

\section{Composite radial structure}

\subsection{Spatial frequency per inner torus radius}\label{sec-sfreq}

The radius $\Rin$ of the inner boundary of the torus dust distribution
is thought to be set by dust sublimation. In this case, for a given
sublimation temperature and grain size distribution, $\Rin$ is
expected to be proportional to $L^{1/2}$, where $L$ is the UV/optical
luminosity of the central engine \citep[e.g.][]{Barvainis87}. This
propotionality was recently confirmed by \cite{Suganuma06} through
near-IR reverberation measurements. These provide the inner torus
radius as the light travel distance for the time-lag between the
variability in the optical and the near-IR.  It is conceivable that
the dust temperature structure within the torus is primarily
determined by the illumination from the central source and is, thus,
also expected to scale with $L^{1/2}$. In this case, if we view
various interferometric size information as a function of spatial
scales normalized by $\Rin$, at least the primary luminosity
dependence of the data, as well as the distance dependence, is
eliminated.

A direct observable in an interferometric measurement is the
visibility, or the normalized Fourier amplitude of the brightness
distribution of a source, as a function of spatial scales (or spatial
wavelength; see below) given in angular size.  Here we normalize the
spatial scale by the angular size of $\Rin$.  More precisely, the
Fourier component is studied as a function of spatial frequency,
i.e. the number of spatial cycles in a given angular size. Thus we
study here visibilities as a function of the number of spatial cycles
in $\Rin$, or spatial frequency per $\Rin$.

For a given projected baseline $b$ in m, observing wavelength
$\lambda$ in $\mu$m, and $\Rin$ in milli-arcsecond (mas), the spatial
freqency in cycles per $\Rin$, written here as $u \cdot \Rin$ (where
$u$ is a spatial frequency per mas), is given as
\begin{equation}
u \cdot \Rin = 4.85 \times 10^{-3} \frac{b\ {\rm (m)} \ }{\lambda\
  {\rm ({\mu}m)}} \cdot \Rin\ {\rm
(mas)} \ \ \ {\rm (cycles\ per}\ \Rin{\rm
)}.
\end{equation}
The corresponding spatial wavelength $\Lambda$, i.e. the reciprocal of
spatial frequency $u$, is given in units of $\Rin$ as
\begin{equation}
\Lambda / \Rin = 2.06 \times 10^{2}\ \frac{\lambda\ {\rm ({\mu}m)}}{b\
    {\rm (m)}} \cdot \frac{1}{\Rin\ {\rm (mas)}}.
\end{equation}
This corresponds to the representative spatial resolution of the
configuration ($b$ and $\lambda$). For simple geometries, the
visibility takes the first null at a spatial wavelength $\Lambda$
roughly equal to the characteristic size (e.g. diameter of an uniform
disk or a ring).  The quantity $\Lambda$ can also be compared directly
to the diffraction limit of $2.5 \times 10^2 \ {\rm (mas)} \ \lambda
{\rm ({\mu}m)} / D {\rm (m)} $ for a single aperture diameter $D$.  We
will study visibilities as a function of these two quantities, $u
\cdot \Rin$ and $\Lambda / \Rin$, throughout this paper.

An accurate scale length for the inner boundary radius $\Rin$ has not
been well determined yet. As the most plausible, observationally
motivated quantity, we adopt the one obtained from the near-IR
reverberation time-lag radii, or, if not available, the overall fit to
them as a function of UV/optical luminosity $L$ given by
\cite{Suganuma06}.  In the latter case, the angular size for the
time-lag radius is given as a function of observed flux $f_{\nu}$ in
$V$-band as (the same as Eq.4 in \citealt{Kishimoto07})
\begin{equation}
\Rin = 0.43 \ [f_{\nu} (V)  \ / \ (50 \ {\rm mJy})]^{1/2} \ {\rm (mas)}.
\label{eq-Rin}
\end{equation}
We note that the time-lag radius is actually a factor of $\sim$3
smaller, for a given luminosity $L$, than the sublimation radius given
by \cite{Barvainis87} assuming graphite grains with radial size
$a$=0.05 $\mu$m and sublimation temperature $\Tsub$=1500 K.  The
possible reasons for this difference have been discussed by
\cite{Kishimoto07}, including the possibility that $\Rin$ is
determined by much larger grains. As we discuss below, the
reverberation radius is supported by the near-IR interferometry of NGC
4151 \citep{Swain03}.  The uncertainty in the time-lag radii is
$\sim$0.2 dex, based on the scatter of the fit.


\subsection{Data}

The data most suitable for the radial structure study are those for
Type 1 AGNs. In these objects, the tori are thought to have much
smaller inclinations than in Type 2 AGNs, so that the radial structure
can be studied without a significant orientation effect. We would also
expect that the dependence on the exact PAs of the projected baselines
is relatively small (the torus image projected on to the sky would not
be far from being circular-symmetric; see more below).  Those Type 1
objects for which long-baseline IR interferometric data are found in
the literature or archive are listed in Table~\ref{tab-obj}, along
with the adopted size for $\Rin$.  All the mid-IR data here were
obtained with VLTI/MIDI.  In addition, we also use the MIDI data for
the Type 2 AGN NGC1068 of \cite{Jaffe04}. The data provide the highest
spatial frequency information in the mid-IR, and the inclination
effect at these wavelengths is expected to be relatively small
compared with that in the near-IR (see more below).

All the MIDI data were uniformly re-reduced with the software EWS
(version 1.5.2; \citealt{Jaffe04SPIE}), with a few modifications using
our own IDL codes. First, since MIDI visibility measurements for the
faint Type 1 targets ($\lesssim$ 1 Jy) are largely limited by the
accuracy of photometric (total flux) measurements, we implemented an
additional background subtraction for the photometry frames.  Second,
to reduce errors in group delay determinations, we have smoothed delay
tracks over $\sim$10-20 frames. To avoid possible positive bias in
correlated flux, we also averaged $\sim$10-20 frames for the
determination of phase offsets.

The system visibility was obtained from the observations of visibility
calibrators usually taken right after or before each target
observation with a similar airmass.  These data were also reduced with
similar smoothings above to calibrate out the effect of the time
averaging.  These calibrators are also generally the photometric
standards found in the list by \cite{Cohen99IR}.  We obtained the
correlated flux, which was corrected for the system visibility, and
total flux separately.  The errors for the total flux and correlated
flux were estimated from the fluctuation of the measurements over
time. For the targets with a few total flux measurements available, we
took weighted means.  The errors for the final visibilities were
obtained from the estimated errors in the correlated flux and total
flux (except for NGC1068 where the errors were estimated from the
scatter of several available visibility measurements).

\subsection{Mid-IR surface brightness distribution}

\begin{figure}
\centering \includegraphics[width=9cm]{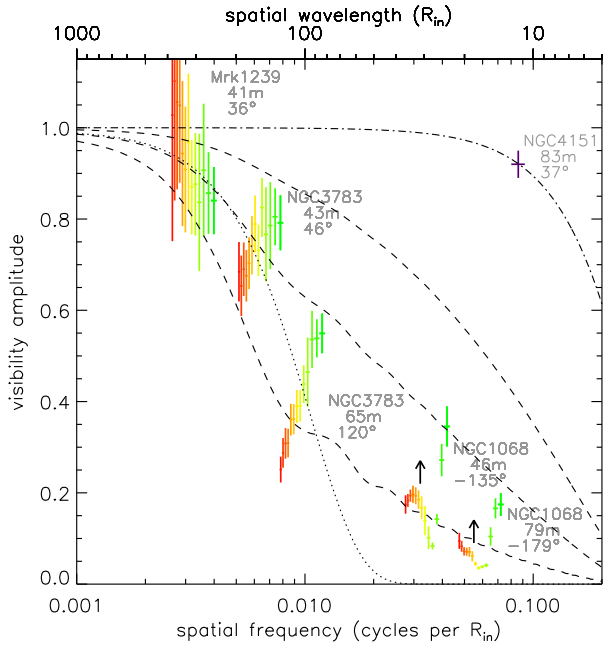}
\caption{Observed visibilities as a function of spatial frequency per
  inner torus radius. The red/yellow/green colors correspond to from
  13 to 8.2 $\mu$m and purple corresponds to 2.2 $\mu$m. The mid-IR
  data are binned with $\Delta \lambda \sim 0.4 \mu$m.  The arrows are
  to show that the NGC1068 data are taken as the lower limit for the
  visibilities of Type 1s at the same spatial frequencies. Visibility
  curves for various simple geometries are also plotted; three dashed
  curves, power-law with index -2.5, -2.0, and -1.5 from top to
  bottom; dotted, Gaussian with HWHM 25 $\Rin$; dot-dashed, ring with
  radius $\Rin$.}
\label{fig-vis-pl}
\end{figure}

Fig.\ref{fig-vis-pl} shows the visibility data listed in
Table~\ref{tab-obj} as a function of spatial frequency per $\Rin$ (or
spatial wavelength $\Lambda$ in units of $\Rin$; upper $x$-axis). The
$x$-axis is in log scale to cover a large range of spatial frequencies
as given by different objects and by different observing wavelengths.
The red/yellow/green colors correspond to wavelenths from 13 to 8.2
$\mu$m, while purple is for 2.2 $\mu$m. The MIDI data have been binned
with $\Delta \lambda \sim 0.4$ $\mu$m, with the error for each bin
taken as the median of the errors over the binned spectral channels
(since errors over adjacent channels are correlated).  For comparison,
we plot the visibility curves for ring, gaussian, and power-law
intensity distributions of different indices with an inner cut-off
radius $\Rin$.

At a given observing wavelength, as the torus inclination increases
from face-on to edge-on cases (i.e. from Type 1 to Type 2
inclinations), the visibilities are expected to generally decrease,
since the inner bright radiation becomes obscured and this effectively
makes the overall size of the intensity distribution larger. (Here we
are restricting our discussion to the spatial frequencies within the
first lobe, i.e. those less than the one at which the visibility
reaches the first null.)  On the other hand, this decrease is expected
to be quite small at long wavelengths, such as in the mid-IR, due to
much smaller obscuration effects.  Therefore, the mid-IR visibilities
of the Type 2 AGN NGC1068 provide quite a meaningful lower limit for
those of the Type 1 objects at the same spatial frequencies (indicated
by the upward arrows in Fig.\ref{fig-vis-pl}).

If we put together the lower limit from NGC1068 with the visibilities
observed for the Type 1 AGNs in the mid-IR (Fig.\ref{fig-vis-pl}), the
data at different spatial freqencies appear quite coherent,
i.e. consistent with the case where these objects share roughly a
common radial brightness distribution in face-on inclinations.  When
compared with various model curves shown in Fig.\ref{fig-vis-pl}, the
mid-IR brightness distribution appears to be consistent with a
power-law form, where the index is $\sim -2$ at $\sim$8 $\mu$m.

The visibility functions for power-law distributions shown in
Fig.\ref{fig-vis-pl} are for the cases with the outer cut-off radius
$\Rout=100\Rin$. For the power-law distributions with index $\ge -2$,
$\Rout$ changes the shape of visibility functions, especially at low
spatial frequencies. Here, the data at high spatial frequencies
(NGC1068 and the longer baseline data for NGC3783) roughly fix the
power law index. Then the low spatial frequency data further constrain
$\Rout$, which corresponds to the outer radius of the mid-IR emission
region, to be roughly $\sim$100$\Rin$ (see also section~3).

In order to be least sensitive to the possible dependency of the
visibilities on the position angle (PA) of the projected baseline, we
would ideally want to compare the visibilities of each object as well
as different objects measured at the major axis direction of the torus
projected on to the sky.  The minor axis PA can be estimated from a
linear radio jet structure ($\sim$77\degr\ for NGC4151,
\citealt{Mundell03}; $\sim$0\degr\ for NGC1068,
e.g. \citealt{Gallimore04}) or optical polarization PA
($\sim$136\degr\ for NGC3783, \citealt{Smith02}; $\sim$40\degr\ for
Mrk1239, \citealt{Goodrich89nls1}). The PA difference between the
projected baseline and the expected major axis is listed in
Table~\ref{tab-obj}. Unfortunately, these relative PAs are not uniform
for the data gathered here. However, they do not seem to disturb the
composite visibility diagram shown in Fig.\ref{fig-vis-pl}, which
might indicate that the PA dependence is small as we expect and
assumed here for Type 1s. Further data are needed to address this
issue.  The two data sets for the Type 2 object NGC1068 have quite
different PAs, which is probably adequate for obtaining lower limits
for Type 1 visibilities.

The only existing near-IR measurement for Type 1 AGNs, i.e. the one
for NGC 4151 (Fig.\ref{fig-vis-pl}, purple cross), is consistent with
a ring of radius $\Rin$.  It is quite expected that the torus shows a
much more compact structure in the near-IR than in the mid-IR, since
the former reflects the distribution of materials at higher
temperatures.  For a given inner boundary radius $\Rin$, the ring
model represents essentially the most compact structure, giving the
upper limit for the visibilities.  This essentially means that $\Rin$
generally cannot be much larger than assumed here (the observed data
points would move to the right in Fig.\ref{fig-vis-pl} if we adopt
larger $\Rin$ for each object, which would exceed the upper limit
given by the $\Rin$ ring). Future near-IR data at high spatial
frequencies are crucial to settle this issue.

Note that we should in principle correct the observed visibility for
the possible contribution from the unresolved accretion disk in the
near-IR to estimate the intrinsic visibility of the torus alone. The
correction depends on the estimate of the near-IR flux fraction from
the disk, which in turn depends on the assumed near-IR spectral shape
of the accretion disk. The correction, however, is estimated to be
very small (less than 5\% reduction; \citealt{Kishimoto07}), as long
as the disk spectrum is bluer than observed in the UV/optical which
seems quite likely (e.g. \citealt{Kishimoto08}).

All the visibility functions shown in Fig.\ref{fig-vis-pl} reach the
first null at a spatial wavelength $\Lambda \sim$ a few $\Rin$
(outside the figure) due to the inner boundary diameter being 2$\Rin$.
In the mid-IR, detailed intensity distributions in the inner few
$\Rin$ region will not affect the overall visibility curves
significantly, since a major fraction of the mid-IR radiation
originates from larger regions.  To discuss whether the same model can
account for both the mid-IR and near-IR data, we need a physical torus
model.

\section{Torus model}

We aim here to generically describe the torus surface brightness
distribution with as a few parameters as possible.  In various torus
models, two types of dust distributions have been considered; namely,
smooth and clumpy distributions
\citep[e.g.][]{Pier92,Pier93,Granato94,Efstathiou95,Nenkova02,Dullemond05,
Schartmann05,Hoenig06,Schartmann08}. We first consider the latter and
assume that the torus consists of various discrete dust clouds each
being optically thick.  The constraints from the alternative case of
smooth dust distribution will be considered later below.

\begin{figure}
\centering 
\includegraphics[width=9cm]{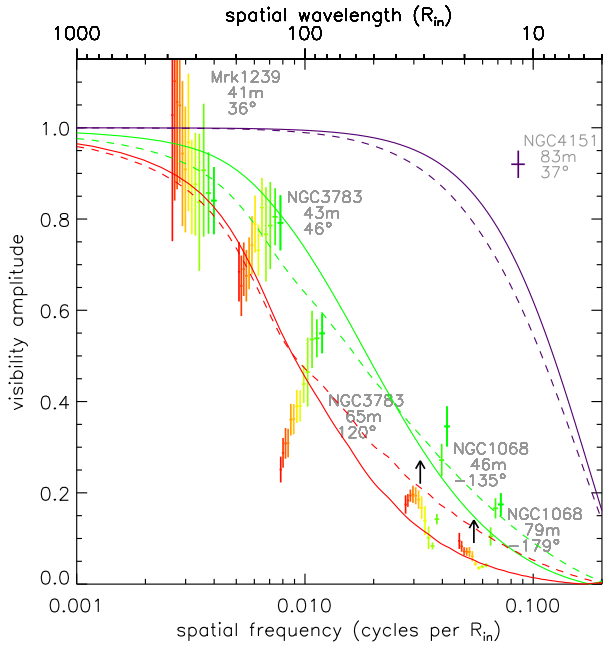}
\caption{Observed visibilities as compared with physical torus
models. Red/green/purple curves are for the observing wavelengh of
13/8.5/ 2.2 $\mu$m , respectively. Solid curves are with $\alpha=0.0$
for the $\gamma=0$ case, while dotted curves are with $\alpha=-0.9$
for the $\gamma=1.6$ case.}
\label{fig-vis-model}
\end{figure}

The near- and mid-IR torus radiation is primarily dominated by that
from those clouds which are directly illuminated by the
central engine (e.g. \citealt{Hoenig06}).  We can then approximately
parameterize the face-on surface brightness distribution $S_{\nu}(r)$
at an observing frequency $\nu$ as a function of radius $r$ as
\begin{equation}
S_{\nu}(r) = I_{\nu}(\Tsf(r)) \cdot ( r / \Rin)^{\alpha}.
\end{equation}
The first term, $I_{\nu}$, is the intensity 
from an illuminated cloud as a function of
$\Tsf(r)$, which is the maximum temperature of the dust grains at
radius $r$ written as
\begin{equation}
\Tsf = \Tsub \cdot ( r / \Rin )^{-\frac{2}{4+\gamma}}.
\end{equation}
This is derived from the thermal equilibrium of a dust grain and
involves the wavelength dependency of the absorption efficiency of the
grain in the form of its spectral index $\gamma$ in the infrared
\citep{Barvainis87}. For the large grain limit, $\gamma$=0, while for
the standard interstellar material (ISM) dust grains, $\gamma \sim
1.6$. The sublimation temperature $\Tsub$ is taken here as 1500K, but
our conclusions below are not sensitive to this exact value of $\Tsub$.

The specific intensity $I_{\nu}$ from an illuminated dust cloud
needs to be obtained with a proper radiative transfer calculation, but
is approximately of the form
\begin{equation}
I_{\nu} = \int e^{-\tau} B_{\nu}(T(\tau)) \ d\tau
\end{equation}
for an optically thick cloud, where $T(\tau)$ is the temperature of
the dust grains at an optical depth $\tau$ from the surface of the
cloud, and $B_{\nu}(T)$ is the Planck function.  At a given observing
frequency, however, we can assume that the intensity distribution over
the torus radius, $I_\nu(r)$, approximately scales with
$B_{\nu}(\Tsf(r))$.

The second term in the expression for $S_{\nu}$, $(r/\Rin)^{\alpha}$,
describes the distribution of the normalized surface density, or the
filling factor per unit area, of the illuminated material as a power
law form.  An implicit assumption here is that the dust cloud
distribution in the innermost region is opaque so that $S_{\nu}(\Rin)
= I_{\nu}(\Tsub)$, and this is the maximum brightness so that $\alpha
\le 0$.

In this way, we can model the visibilities as a function of the
parameter $\alpha$ for an assumed dust grain size distribution fixed
by $\gamma$.  We consider the two cases above; namely, the large grain
limit ($\gamma=0$) and the ISM case ($\gamma = 1.6$).  For each case,
we can estimate an adequate density index $\alpha$ for roughly
reproducing the visibilities observed in the 8-13 $\mu$m range.  We
show in Fig.\ref{fig-vis-model} such representative curves where
$\alpha$ is taken to be 0.0 and -0.9 for the $\gamma = 0$ and 1.6
cases, respectively. We estimate the uncertainties in these indices to
be $\sim$0.3, based on the uncertainty in the visibility measurements
and also in $\Rin$ ($\sim$0.2 dex; see sec.~\ref{sec-sfreq}).

All the curves shown in Fig.\ref{fig-vis-model} have been calculated
with an outer boundary of 100 $\Rin$. The results do not change
significantly if a larger outer boundary is adopted (the region
outside 100$\Rin$ contains only a small fractional amount in the
mid-IR).  Therefore, in the $100\Rin$ cases adopted here, the outer
boundary is set almost by the temperature distribution, rather than
the density distribution of the illuminated material.

Let us consider briefly the case of smooth dust distribution.  If the
opening angle of the geometrical distribution is constant over the
radii, i.e. if there is no geometrical flaring, the material at large
radii is illuminated only by the attenuated radiation from the central
engine. Then $I_{\nu}$ would represent the intensity from the dust
grains at radius $r$ where the distribution of the maximum
temperature $\Tsf(r)$ is steeper\footnotemark[1] (decreasing more
quickly with radius) than considered for the clumpy case
above. Therefore, the normalized surface density distribution,
$(r/\Rin)^{\alpha}$, has to be shallower, closer to the upper limit of
$\alpha = 0$.  If there is a geometrical flaring, then we would revert
to consider the directly illuminated surface at each radius, leading
to essentially the same conclusions obtained for the clumpy case.

\footnotetext[1]{We note that if indirect heating from the re-emission
of dust grains is important, the temperature distribution $\Tsf(r)$
becomes less steep, so that the constraint difference from the
clumpy case will be smaller.}

As for the consistency with the near-IR data, the large grain case
produces slightly more compact distributions in the near-IR than the
ISM case (because the temperature distribution is steeper), and is
marginally consistent with the NGC4151 data. Thus the large grain case
is slightly favored here.  We note that the large grain case is
consistent with the small innermost radii suggested by the near-IR
reverberations, though other contributing factors are not ruled out
\citep{Kishimoto07}.  While we would need a more detailed modeling of
the surface brightness distribution for the innermost region, we also
need more observational data in the near-IR to constrain such detailed
models.

\section{Conclusions}

The long-baseline IR interferometry data for AGN tori are still
generally limited in $uv$ coverage for each target. The data for
various objects probe different spatial scales. We argue that one way
to uniformly study these various data is to view them as a function of
spatial frequency per inner torus boundary radius $\Rin$, or spatial
scale in units of $\Rin$.  In this way, using primarily the data for
Type 1 AGNs, we have tried to construct a composite visibility
function over a wide range of spatial scales and investigated the
radial structure of the tori.  The data obtained so far suggest a
common radial distribution of the face-on surface brightness in the
mid-IR that is approximately of a power-low form with index $\sim -2$
and extends to $\sim$100$\Rin$. Considering the temperature
distribution of the dust grains, this corresponds to the surface
density distribution of the directly illuminated material ranging
approximately between $r^{0}$ and $r^{-1}$.  We aimed here to derive
direct constraints with a simple model.  Further data are definitely
needed over the wavelengths from the near-IR to mid-IR to test the
composite visibility function and to further constrain models.

\begin{acknowledgements}

This research is partly based on observations made with the European
Southern Observatory telescopes obtained from the ESO/ST-ECF Science
Archive Facility.  


\end{acknowledgements}



\end{document}